\documentclass{article}
\usepackage{graphicx,array,cite,dcolumn,amsmath,amssymb,hhline,hyperref,color,float,theorem,mdwlist,cancel,multirow,hyperref}
\usepackage[latin2]{inputenc}
\usepackage[english]{babel}
\newcommand{\brackets}[1]{\left({#1}\right)}

\title{\includegraphics[width=0.35\textwidth]{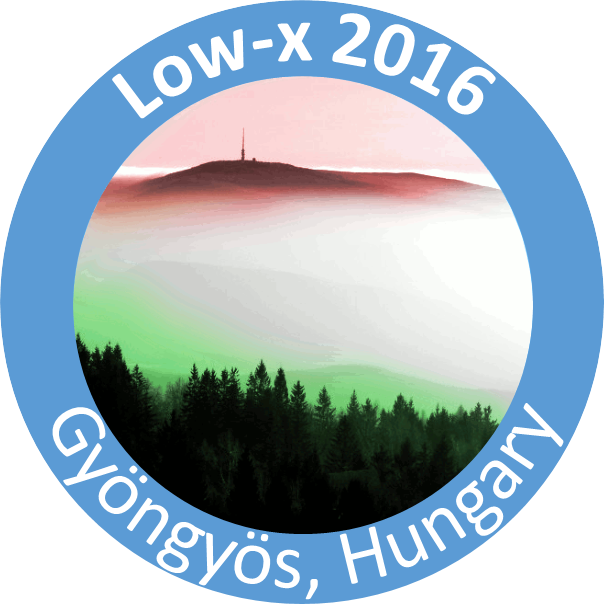}\\[1cm]
Initial energy density of p+p collisions at the LHC}
\author{{M.~Csan\'ad$^1$, T.~Cs\"org\H{o}$^{2,3}$,}\\[1ex]
$^1$E\"otv\"os Lor\'and University, Budapest, Hungary\\
$^2$Wigner Research Centre for Physics, Budapest, Hungary\\
$^3$Eszterh\'azy K\'aroly University KRC, Gy\"ongy\"os, Hungary\\
}

\begin{document}

\fontfamily{lmss}\selectfont
\maketitle

\begin{abstract}
Accelerating, exact, explicit and simple solutions of relativistic hydrodynamics allow for a simple
description of highly relativistic p+p collisions. These solutions yield a finite rapidity distribution,
thus they lead to an advanced estimate of the initial energy density of high energy collisions.
We show that such an advanced estimate yields an initial energy density  in $\sqrt{s}=7$ and 8 TeV p+p
collisions at LHC around or above the critical energy density from lattice QCD, and a corresponding initial temperature
above the critical temperature from QCD and the Hagedorn temperature. We also show, that several times the
critical energy density may have been reached in high multiplicity events, hinting at a non-hadronic medium created in
high multiplicity $\sqrt{s}=7$ and 8 TeV p+p collisions.
\end{abstract}

\section{Introduction}
The interest in relativistic hydrodynamics grew in past years mainly due to the
discovery of the almost perfect fluidity of the experimentally created Quark-Gluon-Plasma (QGP) at
the Relativistic Heavy Ion Collider (RHIC)~\cite{Adcox:2004mh}.
Hydrodynamical models aim to describe the space-time picture of heavy-ion collisions
and infer the relation between experimental observables and the initial conditions.
Besides numerical simulations there is also interest in models where exact solutions
of the hydrodynamical equations are used.
It is customary to describe the medium created in heavy ion collisions 
with hydrodynamic models, however
such applications to p+p and h+p collisions are rare~\cite{Agababyan:1997wd},
because the proton-proton system is frequently
considered as not hot and dense enough to create a supercritical (non-hadronic) medium.
Energy densities in $\sqrt{s}=200$ GeV p+p collisions are definitely below this limit. 
It is however an interesting question, how high energy densities can be
reached in the  $\sqrt{s}=7$ TeV p+p collisions at the LHC. In this paper
we will show how pseudorapidity distributions can be calculated from a hydrodynamic
solution, compared to data, and how this can be used to estimate the initial
energy density of high-energy collisions.

\section{Hydrodynamics}
The basic hydrodynamical equations are the local continuity and energy-momentum-conservation equations:
\begin{align}
\partial_{\nu} (nu^{\nu}) = 0,\qquad
\partial_{\nu}T^{\mu\nu} = 0,
\end{align}
with $n$ being a conserved charge, and $T$ is the energy-momentum tensor. In case of a perfect fluid it is
\begin{align}
T^{\mu\nu}=(\epsilon + p) u^{\mu}u^{\nu}-pg^{\mu\nu}.
\end{align}
where $\epsilon$ is the energy density and $p$ the pressure. The Equation of State (EoS) closes the set of equations:
$\epsilon = \kappa p$ while $p=nT$ defines temperature $T$. An analytic hydrodynamical solution is a functional
form of $\epsilon$, $p$, $T$, $u^\mu$ and $n$, which solves the above equations. 

We discuss the solution detailed in refs.~\cite{Csorgo:2006ax,Csorgo:2007ea,Nagy:2007xn,Csorgo:2008pe}:
\begin{gather}
u^\mu = ({\rm ch}\lambda\eta,{\rm sh}\lambda\eta),\;
n =  n_f\frac{\tau_f^{\lambda}}{\tau^{\lambda}},\;
T =  T_f\left(\frac{\tau_f}{\tau}\right)^{\frac{\lambda}{\kappa}}\label{e:sol3}.
\end{gather}
Here $\tau$ is a coordinate proper-time, $\eta$ the space-time rapidity, subscript $f$ denotes
quantities at the freeze-out, while $\lambda$ controlls the acceleration.
If $\lambda=1$, there is no acceleration and we get back the
accelerationless Bjorken solution of ref.~\cite{Bjorken:1982qr}.

\section{Rapidity distributions}
The differential rapidity distribution or rapidity density $dN/dy$ (with $N$ being then the total number of particles) was calculated
in refs.~\cite{Csorgo:2006ax,Csorgo:2007ea,Nagy:2007xn,Csorgo:2008pe}:
\begin{align}\label{e:dndy-approx}
\frac{dN}{dy}\approx N_0 \cosh^{\frac{\alpha}{2}-1}\brackets{\frac{y}{\alpha}}
e^{-\frac{m}{T_f}\cosh^\alpha\brackets{\frac{y}{\alpha}}} ,
\end{align}
with $\alpha=\frac{2\lambda-1}{\lambda-1}$, and $N_0$ is a normalization parameter.

The rapidity distribution is approximately Gaussian, if $\lambda>1$. At $\lambda=1$, the
districution becomes flat, as this is the Bjorken limit (corresponding to the Hwa-Bjorken solution).
Also note that in order to describe experimental data, pseudorapidity distributions have
to be calculated as well. See details in refs.~\cite{Csorgo:2006ax,Csorgo:2007ea,Nagy:2007xn,Csorgo:2008pe}.


\section{Energy density estimation}
In this section we show how this model can be used for
improving the famous energy density estimation made by Bjorken~\cite{Bjorken:1982qr}. 
We modify Bjorken's method to take into account acceleration
effects. Let us focus on the thin transverse slab at mid-rapidity,
just after thermalization ($\tau=\tau_0$), illustrated by Fig. 2 of ref.~\cite{Bjorken:1982qr}. 
The radius $R$ of this slab is estimated by the
radius of the colliding hadrons or nuclei, and the initial ``fireball'' volume is $dV=(R^2\pi)\tau_0 d\eta_0$,
where $\tau_0 d\eta_0$ is the longitudinal size, as $d\eta_0$ is the pseudorapidity width at $\tau_0$.
See refs.~\cite{Csorgo:2006ax,Csorgo:2007ea,Nagy:2007xn,Csorgo:2008pe} for details.
The energy content is $dE = \langle E\rangle dN$, where $dN$ is the number of particles and $\langle E \rangle$ is their
average energy near $y=0$. So, as given in Bjorken's paper, the initial energy density is
\begin{align}\label{e:Bjorken}
    \epsilon_{\rm Bj} = \frac{\langle E\rangle dN}{(R^2 \pi)\tau_0d\eta_0} =
\frac{\langle E\rangle}{(R^2 \pi)\tau_0}\left.\frac{dN}{d\eta}\;\right|_{\eta=\eta_0} .
\end{align}
Here $\tau_0$ is the proper-time of thermalization, estimated
by Bjorken as $\tau_0\approx 1$fm.

For accelerationless,
boost-invariant Hwa-Bjorken flows $\eta_0=\eta_f=y$, however, for
accelerating solutions one has to apply a correction to take into
account the acceleration effects on the energy density estimation, see
ref.~\cite{Nagy:2007xn} for details.
Thus, for an EoS of $\kappa=1$, the initial energy density is given by a corrected  estimation $\epsilon_{\rm corr}$ as 
\begin{align}\label{e:ncscs}
\epsilon_{\rm corr}=\epsilon_{\rm Bj}\brackets{2\lambda-1}\brackets{\frac{\tau_f}{\tau_0}}^{\lambda-1}
\end{align}
Here $\epsilon_{\rm Bj}$ is the Bjorken estimation, which is recovered if $dN/dy$ is flat (i.e. $\lambda=1$), but for
$\lambda>1$, both correction factors are bigger than 1. Hence the initial energy densities are under-estimated by the Bjorken
formula. In refs.~\cite{Csorgo:2006ax,Csorgo:2007ea,Nagy:2007xn,Csorgo:2008pe} we performed fits to BRAHMS pseudo-rapidity
distributions from ref.~\cite{Bearden:2001qq}, and these fits indicate that $\epsilon_{\rm corr}=8.5-10$ GeV/fm$^3$ in
Au+Au collisions at RHIC.

The above corrections are exact results, that were derived in details for a special equation of state (EoS) of
$\kappa =1$~\cite{Nagy:2007xn}. The correction factors in eq.~(\ref{e:ncscs}) take into account the work
done by the pressure on the surface of a finite and accelerating, hot fireball. However, the relation of the
pressure to the energy density is obviously EoS dependent, and as proposed in refs.~\cite{Csorgo:2006ax,Csorgo:2007ea,Csorgo:2008pe} the effects of
a non-ideal equation of state can be estimated with the following formula:
\begin{align}
\epsilon_{\rm corr} = \epsilon_{\rm Bj}\brackets{2\lambda-1}\brackets{\frac{\tau_f}{\tau_0}}^{\lambda-1}
\brackets{\frac{\tau_f}{\tau_0} }^{(\lambda-1)(1-c_s^2)} \label{e:conjeps} 
\end{align}
This conjecture satisfies several consistency requirements, for example, it goes back to the exact
result of eq.~(\ref{e:ncscs}) in case of a super-hard EoS of  $c_s = 1$ and gives
initial energy density values that were checked against numerical
solutions~\cite{Csorgo:2006ax}.

From basic considerations~\cite{Bjorken:1982qr}, as well as from lattice QCD calculations~\cite{Borsanyi:2010cj},
it follows that the critical energy density, needed to form a non-hadronic medium is around 1 GeV/fm$^3$. From the
lattice QCD calculations one gets $\epsilon_{\rm crit}=(6-8)\times T_{\rm crit}^4$ (in $\hbar c=1$ units), and
even with a conservative estimate of $T_{\rm crit} = 170$ MeV, one gets $\epsilon_{\rm crit}\lesssim 1$ GeV/fm$^3$.
Thus energy densities above this value of $\epsilon_{\rm crit}\approx 1$ GeV/fm$^3$ indicate the
formation of a non-hadronic medium.

\section{Initial energy density in LHC p+p collisions.}
Let us estimate the quantities in eq.~(\ref{e:Bjorken}).
The average transverse momentum in $\sqrt{s}=7$ TeV p+p collisions is 
$\langle p_t\rangle = 0.545 \pm 0.005_\textnormal{stat}\pm 0.015_\textnormal{syst}$ GeV$/c$~\cite{Khachatryan:2010us},
which corresponds to  $\langle E\rangle=0.562$ GeV$/c^2$ at midrapidity
(assuming most of these particles are pions). The radius $R$ can be
estimated from the elastic and total cross-sections via
$R^2\pi = \sigma_{\rm tot}^2/4\sigma_{\rm el}$. These are measured by TOTEM,
$\sigma_{\rm tot}=98.0\pm2.5$ mb and $\sigma_{\rm el}=25.1\pm1.1$ mb~\cite{Antchev:2013iaa},
From this, $R=1.76\pm0.02$ fm can be estimated.
This is also approximately verified by HBT measurements~\cite{Aamodt:2011kd,Khachatryan:2010un}. The
formation time, $\tau_0$, is conservatively assumed to be 1 fm$/c$. The only remaining parameter is 
 the rapidity density at midrapidity. As measured by the LHC experiments,
the charged particle multiplicity is found to be $6.01\pm0.01\textnormal{(stat)}^{+0.20}_{-0.12}\textnormal{(syst)}$
at ALICE~\cite{Khachatryan:2010us}, while
$5.78\pm0.01_\textnormal{stat}\pm0.23_\textnormal{syst}$ at CMS~\cite{Aamodt:2010pp},
but in some multiplicity classes it may reach values of 25-30
(see table I. of ref.~\cite{Aamodt:2011kd}). We will take the average of the first two values.
The total multiplicity is then $3/2\times$ the charged particle multiplicity.
Based on eq.~(\ref{e:Bjorken}) one gets:
\begin{align}
    \epsilon_{\rm Bj}(7 TeV) = \frac{0.562\times 1.5 \times 5.895}{1.76^2\pi}\;\rm{GeV/fm}^3 = 0.507\;\rm{GeV/fm}^3.
\end{align}

\begin{figure}
\begin{center}
\includegraphics[width=1\linewidth,clip]{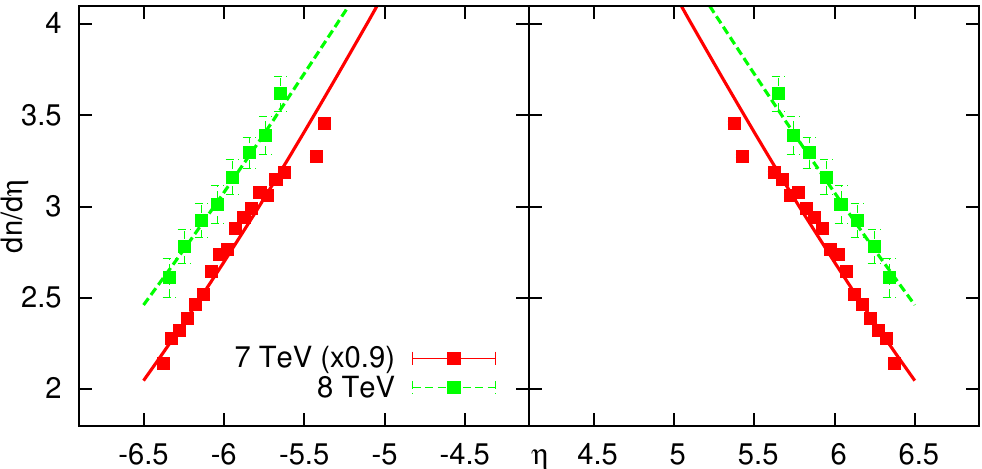}
\end{center}
\caption{\label{f:totemfit} Charged particle
$\frac{dN}{d\eta}$ distributions from TOTEM
fitted with the result of the relativistic hydro solution described
in this paper.}
\end{figure}

The advanced estimate is based on TOTEM pseudorapidity density $dN/d\eta$ data, as these reach
out to large enough $\eta$ values so that the acceleration parameter can be determined. Fits to TOTEM data were
performed via eq.~(\ref{e:dndy-approx}) , as shown in Fig.~\ref{f:totemfit}. The fit resulted in the acceleration parameter
$\lambda = 1.073\pm0.001_\textnormal{stat}\pm0.003_\textnormal{syst}$,
where the systematic error is based on the point-to-point systematic error of the data points.

\begin{figure}
\begin{center}
\includegraphics[width=0.95\linewidth,clip]{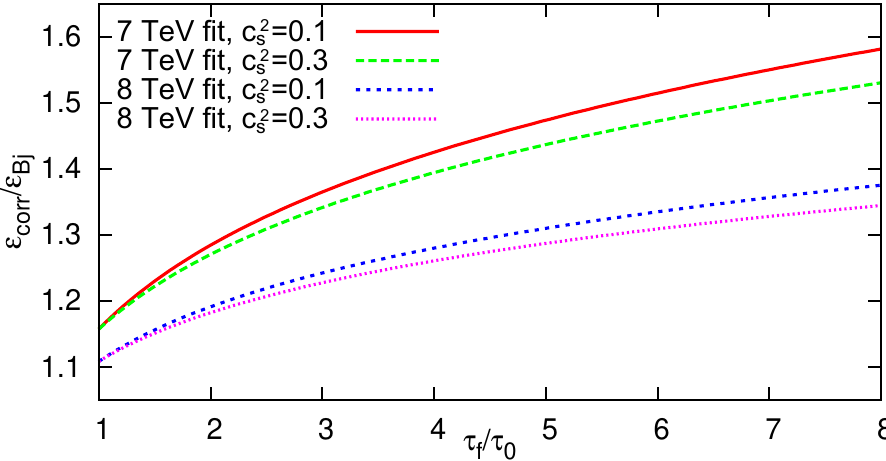}
\end{center}
\caption{\label{f:cscorr} The correction factor as a function of
freeze-out time versus thermalization time ($\tau_f/\tau_0$). At a
reasonable value of ~2, the correction factor is around 25\%.}
\end{figure}

\begin{figure}
\begin{center}
\includegraphics[width=0.99\linewidth]{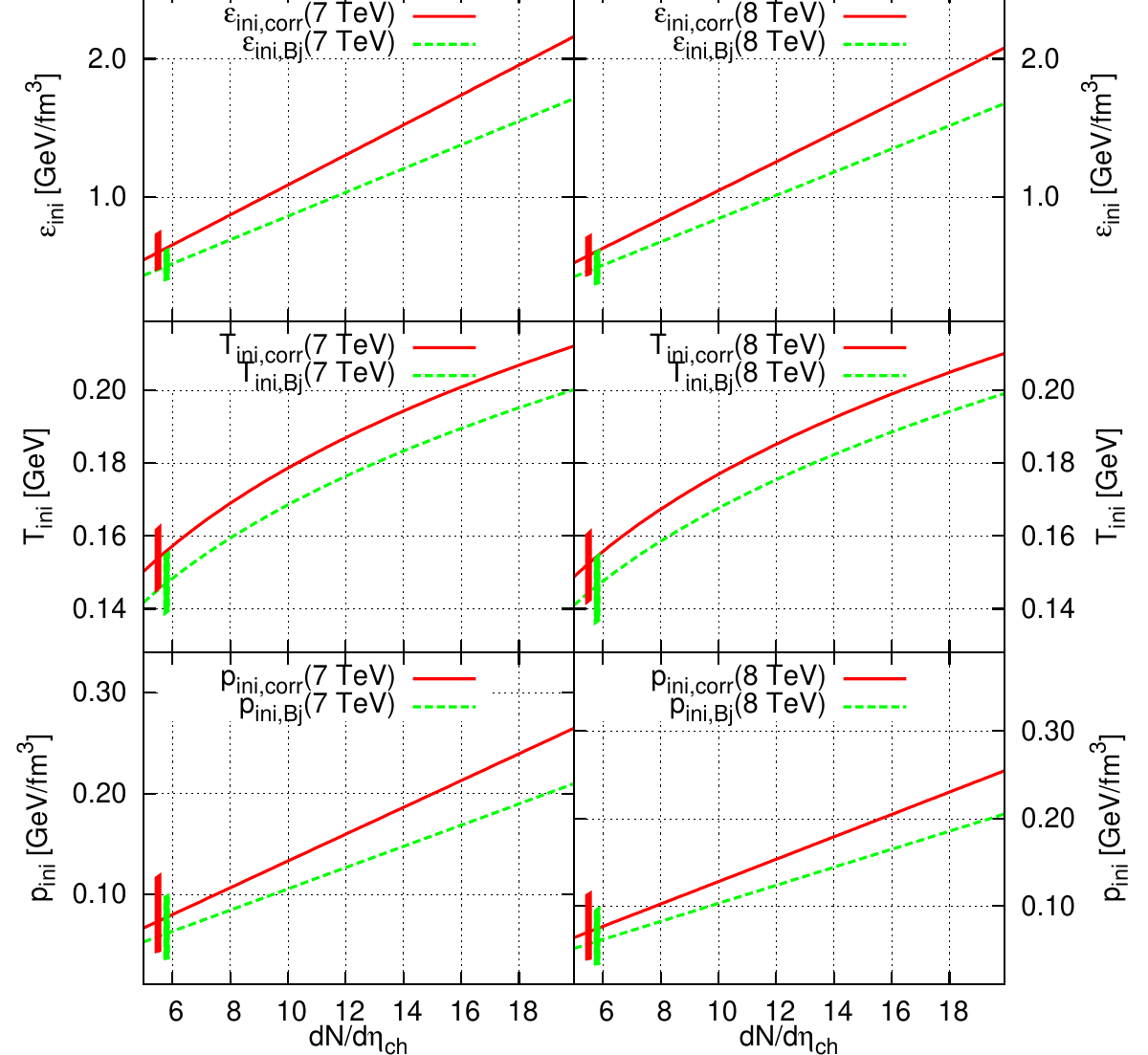}\vspace{-10pt}
\end{center}
\caption{\label{f:multdep}Initial energy density (based on table~\ref{t:errors}), temperature
and pressure (based on the $\epsilon\propto T^4$ relationship) at 7 TeV,
is indicated as a function of central multiplicity density.
The Bjorken-estimate (dashed curve) is above the critical energy density of 1 GeV/fm$^3$ if the
multiplicity is larger than 6-7. Corrected initial energy density (solid curve) is always above
the critical value. Boxes (parallelograms) show systematic uncertainty. For 8 TeV, as noted in the text,
$\epsilon_{\rm Bj}$ increases by $\approx$2\% -- for the sake of clarity we did not plot these
curves.}
\end{figure}

Assuming $c_s^2=0.1$ (this is a quite realistic value, at least no harder EoS is expected at LHC,
as similar EoS was found at RHIC as well~\cite{Lacey:2006bc,Borsanyi:2010cj,Csanad:2011jq}), one only needs a $\tau_f$ value.
As shown in eq.~(\ref{e:sol3}), temperature is proportional to $\tau^{-\lambda/\kappa}$.
From this, $\tau_0 = \tau_f (T_f/T_0)^{\kappa/\lambda}$. Thus if the freeze-out temperature (assumed
to be around the Hagedorn-temperature or the critical temperature of lattice QCD) is $T_f=140$ MeV,
then an initial temperature of $T_0 = 170$ MeV (needed in order to form a strongly interacting quark
gluon plasma) corresponds to $\tau_f$ being 5-6 times $\tau_0$, for $c_s^2=0.1$
and $\lambda=1.1$. Even if $c_s^2$ and $\lambda$ are higher, $\tau_f/\tau_0\approx 5-6$ seems to
be a rather conservative value. With this, one gets the multiplicative correction
factors of 1.146 and 1.101, thus
\begin{align}\label{e:corrfact}
\epsilon_{\rm corr}(7\;{\rm TeV}) = 1.262\epsilon_{\rm Bj}(7\;{\rm TeV})=0.640\;\rm{GeV/fm}^3,
\end{align}
which is below the critical value. The $c_s^2$ and $\tau_f/\tau_i$ dependence of the correction factor
is shown in Fig.~\ref{f:cscorr}. Note, that the average p+p multiplicity was used here, so
this value represents an average energy density in p+p collisions at 7 TeV LHC energy. Based on Table 1 of
ref.~\cite{Aamodt:2011kd}, much larger multiplicities have been reached in the same reaction,
in certain event classes. The energy
density results for these multiplicities is shown on Fig. \ref{f:multdep}. It is clear from this
plot, that even for the original Bjorken estimate, supercritical enegy densities may have
been reached in high multiplicity events. The corrected estimate gives supercritical values
even for somewhat more than average multiplicities. We also calculated the initial temperature based
on the  $\epsilon\propto T^4$ relationship, assuming that 175 MeV corresponds to 1 GeV/fm$^3$
approximately. This is also shown in  Fig.~\ref{f:multdep}, as well as the reachable pressure values. 
An initial temperature of 300-600 MeV may have been reached in 200 GeV central Au+Au collisions of
RHIC~\cite{Adare:2008ab}. Initial temperature values in 7 TeV p+p seem to be lower than that, but
300 MeV can be reached in events with a multiplicity of 45. However, supercritical
initial temperatures of 200 MeV may already be
reached in events with a multiplicity of 10.

Finally let us estimate what happens at $\sqrt{s}=8$ TeV. As for the Bjorken-estimate, we need the change
in charged particle multiplicity, average transverse energy and transverse size. The $s$ dependence of $dN/d\eta$ is
estimated as  $0.715\cdot(\sqrt{s})^{0.23}$ in Ref.~\cite{CMS:2013yca}, this means a 3.12\% increase from 7 TeV to 8 TeV.
Average transverse momentum $s$ dependence is estimated as $\langle p_t \rangle =0.413 - 0.0171 \ln s + 0.00143 \ln^2 s$
in Ref.~\cite{Khachatryan:2010us}, which means a 1.53\% increase in $\langle E \rangle$. Transverse size increase
can be calculated based on cross-section measurements of Refs.~\cite{Antchev:2013iaa,Antchev:2013paa}, 
and one gets a 2.47\% increase in cross-section. Based on Eq.~(\ref{e:Bjorken}), this alltogether means a
2.18\% increase, i.e.\ $\epsilon_{\rm Bj}(8\;{\rm TeV})=0.519$ GeV/fm$^3$.
We also fitted $dN/d\eta$ data from TOTEM~\cite{Chatrchyan:2014qka}
as shown in Fig.~\ref{f:totemfit}. We obtained $\lambda = 1.067\pm 0.001$ in this case. This corresponds to 
a correction factor of 1.240, similarly to Eq.~(\ref{e:corrfact}). Finally, we get
\begin{align}\label{e:8tev}
\epsilon_{\rm corr}(8\; {\rm TeV}) = 1.240\epsilon_{\rm Bj}(8\;{\rm TeV})=0.644\;\rm{GeV/fm}^3.
\end{align}
This value is based on the average multiplicity in $\sqrt{s}=8$ TeV collisions. However, at a fixed multiplicity, there is almost
no difference between the two collision energies: average transverse energy increases by 1.5\%, but cross section also
increases by 2.5\%. This means a 1\% decrease, which is much smaller than the systematic uncertainties in this estimate --
to be discussed in the next section. Fig.~\ref{f:multdep} indicates the multiplicity dependence of $\epsilon_{\rm ini}$,
$T_{\rm ini}$ and $p_{\rm ini}$ for both 7 and 8 TeV p+p collisions.

\section{Uncertainty of the estimate}
Different sources of uncertainties are detailed in Table~\ref{t:errors}. The most
important one comes from $dN/d\eta$ at midrapidity. From fig.~\ref{f:multdep} it is clear that for the
Bjorken-estimate, energy density is above the critical value of 1 GeV/fm$^3$ if the multiplicity is larger than 6-7,
while the corrected initial energy density is always above the critical value. Taking all sources of uncertainties
into account, the final result for the energy density corresponding to mean multiplicity density at 7 TeV is
\begin{align}
\epsilon_{\rm corr}(7\; {\rm TeV})  = 0.64 \pm 0.01\rm{(stat)}^{+0.14}_{-0.10}\rm{(syst)}\;\rm{GeV}/\rm{fm}^3
\end{align}
and the main systematic error comes from the estimation of the ratio $\tau_f/\tau_0$. In the 8 TeV case,
the estimate yields a somewhat larger number (0.644 versus 0.640), but the uncertainties are higher
due to additional uncertainties of extrapolations to 8 TeV.

An important source of systematic uncertainty is the use of the given hydrodynamic solution. This uncertainty
may be estimated by using other hydrodynamic models that contain acceleration: the Landau model~\cite{Landau:1953gs}, the Bialas-Peschanski
model~\cite{Bialas:2007iu}, or numeric models of hydrodynamics,
however, in the current paper we focus on the analytic
results that can improve on Bjorken's famous initial energy density
estimate. A more detailed numerical hydrodynamical investigation is
outside the scope of the present manuscript.

\begin{table}
\begin{center}
\begin{tabular}{|c|c|c|c|}
\hline
  parameter                             &    value                  & stat. & syst. eff. on $\epsilon$ \\\hline
  $\lambda$                             &  1.073                   & 0.1\%                         & 0.4\% (from data) \\\hline
  \multirow{2}{*}{$c_s^2$}   &   0.1                      &  -                                 & -2\%+0.2\%\\
                                             &                              &                                      &(if $0.05<c_s^2<0.5$)\\\hline
  \multirow{2}{*}{$\tau_f/\tau_0$}  &   2                 &  -                                 & -4\%+10\%\\
                                             &                              &                                      &(for $\tau_f/\tau_0$ in 1.5--4) \\\hline
   $\tau_0$  [fm$/c$]              & 1                           & -                                  & underestimates $\epsilon$ \\\hline
   $R$   [fm]                          & 1.76                   & 0.5\%                       & 1.3\% (from $\sigma_{\rm el,tot}$)\\\hline
  $\langle E \rangle$ [GeV$/c^2$]  &  0.562          & 0.5\%                         & 3\%\\\hline
  $dN/d\eta$   (7 TeV)              & 5.895                     & 0.2\%                         & 3\%\\\hline
  \end{tabular}
  \caption{Sources of statistical and systematic errors for the 7 TeV estimate.\label{t:errors}}
\end{center}
\end{table}

\section{Summary}
We have shown, that based on an accelerating solutions of relativistic hydrodynamics and TOTEM LHC data, an advanced
estimate of the initial energy density yields a value that is somewhat below the values
expected for a supercritical state. As the energy density is proportional to the measured multiplicity, in 
high-multiplicity 7 and 8 TeV proton-proton collisions, energy densities several times the
critical energy density of 1 GeV/fm$^3$ have been reached. 
This result means, that an important and necessary condition is satisfied for the
formation of a non-hadronic medium in 7 and 8 TeV p+p collisions at CERN LHC,
however, the exploration of additional signatures (radial and elliptic
flow, volume or mean multiplicity dependence of the signatures of the
nearly perfect fluid in p+p collisions, scaling of the HBT radii with
transverse mass, and  possible direct photon signal and low-mass dilepton enhancement)
should be a subject of detailed experimental investigation even in p+p
collisions at the LHC.

The application of hydrodynamical expansion to data analysis in high
energy p+p collisions is not an unprecedented or new idea, as Landau worked out hydrodynamics for p+p
collisions~\cite{Belenkij:1956cd}, and Bjorken also notes this possibility in his paper~\cite{Bjorken:1982qr}
describing his energy density estimate.

It is also noteworthy that Hama and Padula assumed~\cite{Hama:1987xv} the formation of an
ideal fluid of massless quarks and gluons in
p+p collisions at CERN ISR energies of $\sqrt{s}$ = 53- 126 GeV.
Alexopoulos et al. used Bjorken's estimate to determine the initial energy
density of $\sim1.1\pm0.2$ GeV/fm$^3$ at the Tevatron in $\sqrt{s}$ = 1.8 TeV
p+$\overline{{\rm p}}$ collisions in the E735 experiment~\cite{Alexopoulos:2002eh}, while Lévai and
Müller argued~\cite{Levai:1991be}, that the transverse
momentum spectra of pions and baryons indicate the creation of a
fluid-like quark-gluon plasma in the same experiment at the same Tevatron
energies. However, these earlier works considered the quark-gluon
plasma
as an ideal gas of massless quarks and gluons, while the RHIC
experiments pointed to a nearly perfect fluid of quarks where the speed of
sound is measured to be $c_s \approx 0.35 \pm 0.05$ that is significantly
different from that of a massless ideal gas of quarks and gluons,
characterized by a $c_s = 1/\sqrt{3} \approx 0.57$. Recently, Shuryak
and Zahed also proposed~\cite{Shuryak:2013ke} the application of hydrodynamics for high
multiplicity p+p and p+A collisions at CERN LHC.

The main result of our study indicates, that the initial energy density is apparently large
enough in average or even low multiplicity p+p collisions at the $\sqrt{s}$ = 7 and 8 TeV
LHC energies to create a strongly interacting quark-gluon plasma, so a
smooth evolution with increasing multiplicity is expected, as far as
hydrodynamical phenomena are considered.

Probably the most important implication of our study is the need for an e+p
and e+A collider: as far as we know only in lepton induced proton and heavy
ion reactions can one be certain that a hydrodynamically evolving medium
is not created even at the TeV energy range. The results of lepton-hadron
and lepton-nucleus interactions thus will define very clearly the particle
physics background to possible collective effects. For example, recently
azimuthal correlations were observed in high multiplicity p+p and p+A
as well as in heavy ion reactions (the ridge effect~\cite{CMS:2012qk,Padula:2011yk}), whose origin is
currently not entirely clear. If such a ridge effect appears also in e+p
and e+A collisions, then most likely this effect is not of a hydrodynamical
origin, while if it does not appear in e+p and e+A collisions in the same
multiplicity range as in p+p and p+A reactions, than the ridge is more
likely a hydrodynamical effect.

If indeed a strongly interacting non-hadronic medium is formed in high multiplicity p+p collisions,
than purely the jet suppression in heavy ion collisions does not reveal the true
nature of these systems: the proper measure would be energy loss per unit length (as proposed in ref.~\cite{Csorgo:2009wc}),
which may be quite similar in these systems, even if the total suppression is different.

We are looking forward to measurements unveiling the nature of
the matter created in proton-proton collisions. In experimental p+p data, one should look for
the enhancement of the photon to pion ratio in high multiplicity events (as compared to low multiplicity
ones)~\cite{Tannenbaum:2010yx}, for a hydrodynamic scaling of Bose-Einstein correlation radii or that of
azimuthal asymmetry~\cite{Adare:2006ti}, or even the enhancement of low mass dileptons~\cite{Afanasiev:2007aa}.

\section{Acknowledgments}
The authors thank the important discussions to Simone Giani, Paolo Guibellino, Federico Antinori, Michael Tannenbaum,
Péter Lévai, Sandra S. Padula and the late Endel Lippmaa. We acknowledge the support of the OTKA grant NK101438
and the NKFIH grant FK123842. T. Cs. gratefully acknowledges
partial support from EFOP 3.6.1. M. Cs. was supported by the J\'anos Bolyai Research Scholarship
of the Hungarian Academy of Sciences, and the \'UNKP-17-4 New National Excellence Program of the 
Hungarian Ministry of Human Capacities.

\end{document}